\documentclass[preprint,prb,showpacs]{revtex4}
\usepackage{graphicx}
\usepackage{dcolumn}
\usepackage{bm}
\usepackage{natbib}

\begin{document}

\title{Temperature-dependent excitonic superfluid plasma frequency evolution in a strong excitonic insulator candidate, Ta$_2$NiSe$_5$}

\author{Yu-Seong Seo$^1$} \author{Man Jin Eom$^{2}$} \author{Jun Sung Kim$^{2}$} \author{Chang-Jong Kang$^{2}$} \author{Byung Il Min$^{2}$} \author{Jungseek Hwang$^{1}$}
\email{Electronic address: jungseek@skku.edu}
\affiliation{$^{1}$Department of Physics, Sungkyunkwan University, Suwon, Gyeonggi-do 16419, Republic of Korea\\ $^{2}$Department of Physics, Pohang University of Science and Technology, Pohang 37673, Republic of Korea}

\date{\today}

\begin{abstract}

We investigate a strong candidate (Ta$_{2}$NiSe$_{5}$) of excitonic insulators, using optical spectroscopy. Ta$_{2}$NiSe$_{5}$ has quasi-one dimensional chains along the $a$-axis. We have obtained anisotropic optical properties of a single crystal Ta$_{2}$NiSe$_{5}$ along the $a$- and $c$-axes. The measured $a$- and $c$-axis optical conductivities exhibit large anisotropic electronic and phononic properties. With regard to the $a$-axis optical conductivity, a sharp peak near 3050 cm$^{-1}$ at 9 K, with a well-defined optical gap ($\Delta^{EI} \simeq$ 1800 cm$^{-1}$) and a strong temperature-dependence, is observed. With an increase in temperature, this peak broadens and the optical energy gap closes around $\sim$325 K ($T_c^{EI}$). The spectral weight redistribution with respect to the frequency and temperature indicates that the normalized optical energy gap ($\Delta^{EI}(T)/\Delta^{EI}(0)$) is $1-(T/T_c^{EI})^2$. The temperature-dependent superfluid plasma frequency of the excitonic condensation in Ta$_{2}$NiSe$_{5}$ has been determined from measured optical data.

\end{abstract}

\pacs{78.20.-e, 78.20.Ci, 71.35.-y}

\maketitle

\section{Introduction}

Excitonic insulators (EI), proposed in the 1960s\cite{Mott:1961,Knox:1963,Jerome:1967}, are novel materials exhibiting correlated electronic phases and have attracted the interest of several experimental and theoretical condensed matter physics groups. An EI has a condensation phase of excitons (or electron-hole pairs) as its ground state for a specific condition ($E_{b}>E_{g}$)\cite{Jerome:1967}, where $E_{b}$ and $E_{g}$ are the binding energy and bandgap, respectively. In the EI phase, superfluidity of neutral electron-hole pairs occurs\cite{kozlov:1965}. EI systems are either semiconductors with small bandgaps or semi-metals with small overlaps between the conduction and valence bands\cite{Jerome:1967}. The excitonic condensation in semiconductors occurs through a Bose-Einstein condensation process, while the condensation in semi-metals occurs through the Bardeen-Cooper-Schrieffer (BCS) process\cite{Phan:2010,Zenker:2012}. Typically, chalcogenide compounds are known to form one group of EIs. Several studies have been performed on Ta$_{2}$NiSe$_{5}$, which is one of transition metal chalcogenides\cite{Sunshine:1985,Disalvo:1986,Wakisaka:2009,Wakisaka:2012,Kaneko:2012,Kaneko:2013,Seki:2014, kim:2016,lu:2017,Larkin:2017}. Ta$_{2}$NiSe$_{5}$ has quasi-one dimensional chains along the $a$-axis\cite{Sunshine:1985,Disalvo:1986}. An angle-resolve photoemission (ARPES) study on Ta$_{2}$NiSe$_{5}$ showed that the top of the valence band at the $\Gamma$-point flattened at temperatures below its structural transition temperature ($T_c^{Str}$ = 325 K), and this flat band was interpreted as an excitonic insulating ground state of condensed electron-hole pairs of Ta 5$d$-electrons and Ni 3$d$- and Se 4$p$-holes\cite{Wakisaka:2009}. A recent study also shows that Ta$_{2}$NiSe$_{5}$ is a zero-gap semiconductor, with a transitions to an EI occurring near 326 K (referred as the onset temperature ($T_c^{EI}$))\cite{lu:2017}. There was another very recent ellipsometry spectroscopic study on Ta$_{2}$NiSe$_{5}$\cite{Larkin:2017}; the authors claimed that exciton-phonon complexes in Ta$_2$NiS$_5$ and Ta$_2$NiSe$_5$ are confirmed and their observation agrees with the hypothesis of an excitonic insulator ground state. In this article, we provide a new set of anisotropic optical data of Ta$_{2}$NiSe$_{5}$ obtained using a different optical spectroscopy technique from the ellipsometry technique. We observed the temperature-dependent evolution of the excitonic condensation gap of Ta$_{2}$NiSe$_{5}$. Furthermore, we extracted a very important physical quantity, the excitonic superfluid plasma frequency, of Ta$_{2}$NiSe$_{5}$ from the measured optical data.

\section{Experiments and anisotropic reflectance spectra}

We present the temperature-dependent anisotropic optical properties of Ta$_{2}$NiSe$_{5}$, recorded along two different crystal orientations ($a$- and $c$-axis) using a conventional optical spectroscopic technique. The high-quality single crystal of Ta$_{2}$NiSe$_{5}$ sample was grown by a chemical vapor transport method. Our optical study was performed on a single crystal Ta$_{2}$NiSe$_{5}$ with an area of 2$\times$2 mm$^2$ and a thickness of 0.5 mm. A commercial FTIR-type spectrometer (Vertex 80v, Bruker) and a continuous flow liquid helium cryostat were used to obtain $a$- and $c$-axis reflectance spectra over a wide spectral range (80 - 20,000 cm$^{-1}$) at various selected temperatures between 9 and 350 K above and below the transition temperature ($T_c^{EI}$). We used linear polarized beam to obtain anisotropic optical spectra with an incident angle on the sample of $\sim$10$^{\circ}$. We also used an {\it in-situ} metallization method to obtain accurate reflectance spectra\cite{Homes:1993}. In this method we used the coated 200 nm thick gold for mid- and far-infrared (or aluminum for near-infrared and visible) film on the sample as the reference reflectance. Furthermore, we corrected the measured reflectance with respect to the gold (or aluminum) film by multiplying the absolute reflectance of the gold (or aluminum).

Fig. 1(A) and 1(B) show the measured reflectance spectra of a single crystal Ta$_{2}$NiSe$_{5}$ along the $a$- and $c$-axis respectively. There was a significant difference in the electronic and phononic properties along the two different crystal axis orientations ($a$- and $c$-axes) as we expected. For Ta$_{2}$NiSe$_{5}$, the quasi-one dimensional chains are along the $a$-axis. Several sharp peaks were observed in the $a$-axis reflectance spectra at low temperatures, with similar strong temperature-dependent behaviors. The reflectance below $\sim$2700 cm$^{-1}$ increases gradually with temperature and above the frequency a peak centered near 3200 cm$^{-1}$ grows with lowering the temperature. This behavior is a typical signature of optical gap formation. The phonon modes seem to be screened at high temperatures above 300 K. The inset in Fig. 1(A) depicts the crystal structure of Ta$_{2}$NiSe$_{5}$, which has a layered structure with the $b$-axis as the stack axis. In contrast to the $a$-axis reflectance spectra, the $c$-axis reflectance spectra displays a rather monotonic temperature-dependence, with the reflectance being gradually suppressed over a wide spectral range from 80 to $\sim$15,000 cm$^{-1}$ with the lowering of the temperature. A set of peaks with weak intensity and narrow spectral widths appear at low temperatures. The physical origin of these new set of peaks are not clear yet. The experimentally measured dc resistivity of the $ac$-plane of Ta$_{2}$NiSe$_{5}$ is displayed in the inset of Fig. 1(B). An anomaly in the dc resistivity associated with the structural phase transition temperature ($T_c^{Str}$) was observed close to 322 K\cite{Disalvo:1986}. Below the $T_c^{Str}$, we also observed the splitting of a phonon mode centered around 160 cm$^{-1}$.

\begin{figure}[!htbp]
  \vspace*{-0.2 cm}%
 \centerline{\includegraphics[width=4.2 in]{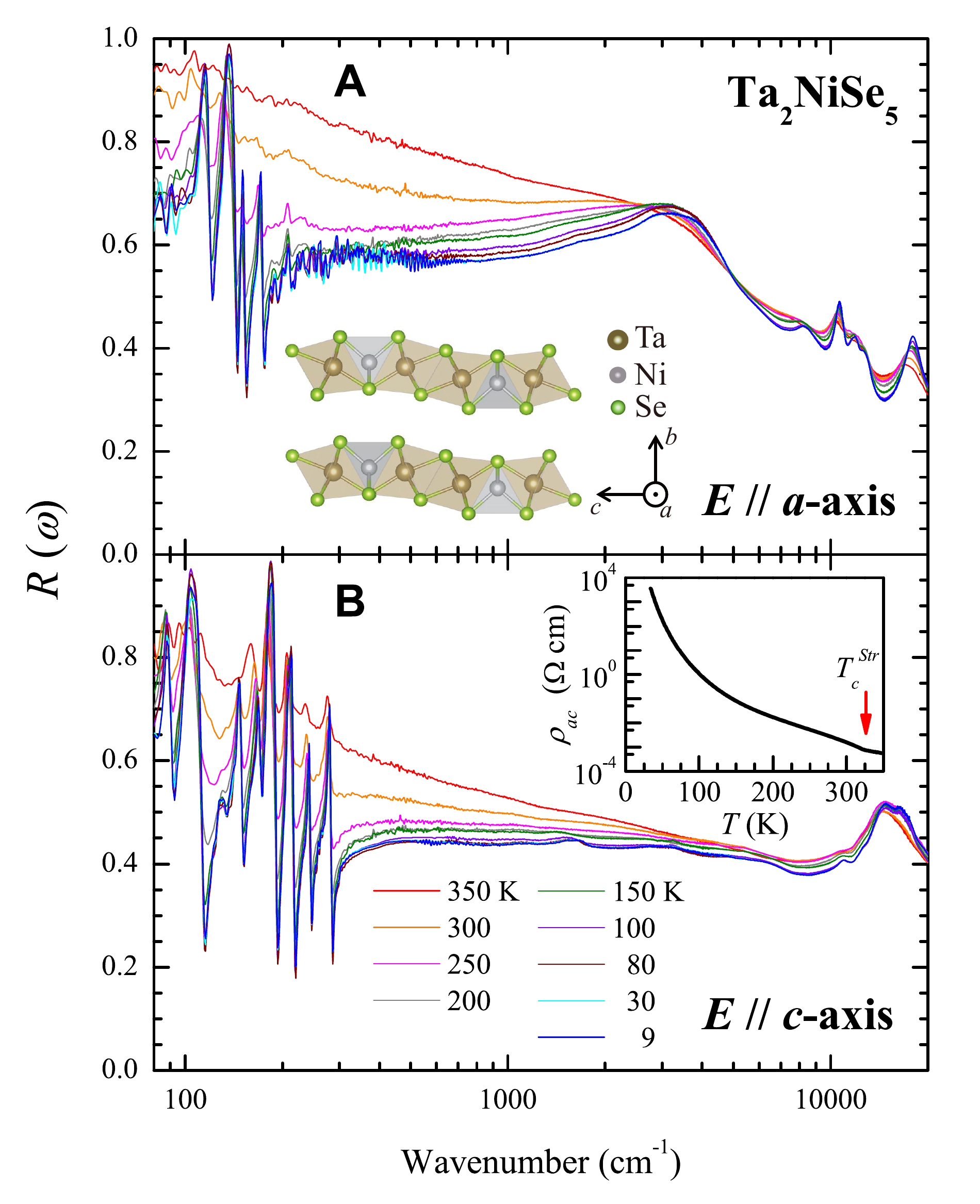}}%
  \vspace*{-0.5 cm}%
\caption{(Color online) Anisotropic reflectance spectra of Ta$_{2}$NiSe$_{5}$ along $a$- and $c$- axes. The measured reflectance spectra of Ta$_{2}$NiSe$_{5}$ along the $a$- and $c$-axes are displayed in (A) and (B) respectively. The spectra were recorded at selected temperatures ranging from 9 to 350 K. The crystal structure of Ta$_{2}$NiSe$_{5}$ is shown in the inset of Fig. 1(A), while the measured dc resistivity of $ac$-plane of Ta$_{2}$NiSe$_{5}$ is depicted in the inset of Fig. 1(B).}
\label{fig0}
\end{figure}

\section{Anisotropic optical conductivity}

\begin{figure}[!htbp]
  \vspace*{-0.3 cm}%
  \centerline{\includegraphics[width=4.2 in]{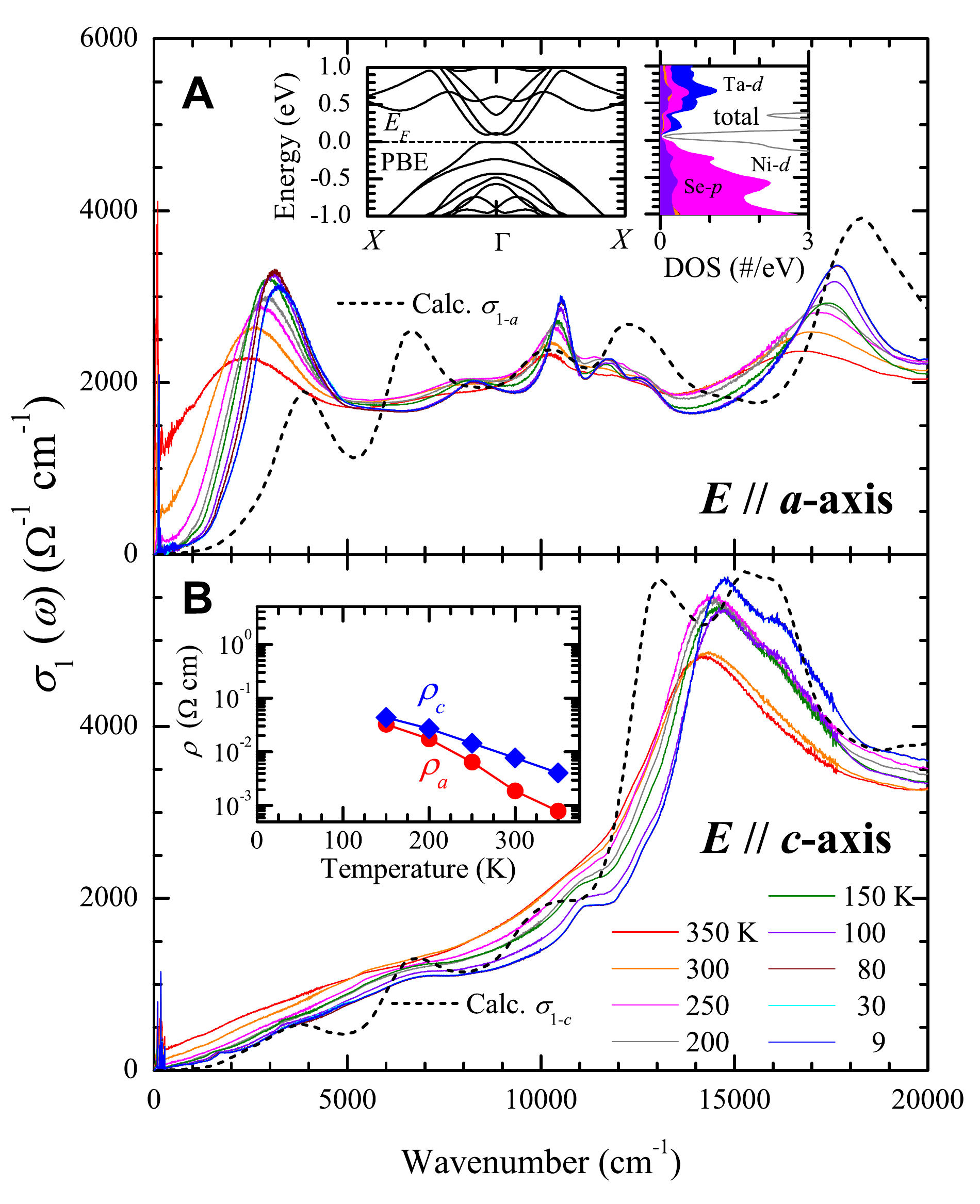}}%
  \vspace*{-0.5 cm}%
\caption{(Color online) Anisotropic optical conductivity of Ta$_{2}$NiSe$_{5}$ along $a$- and $c$- axes. The experimentally optical conductivities of Ta$_{2}$NiSe$_{5}$ along $a$- and $c$-axes are depicted in (A) and (B), respectively. The optical conductivities were obtained with a Kramers-Kronig analysis on the measured reflectance spectra for the different temperatures (within the range of 9 to 350 K). The inset of Fig. 2(A) shows the band dispersion diagrams obtained via the PBE functional and the partial densities of states, which were obtained from the first-principles calculations. The inset in Fig. 2(B), shows the extracted dc resistivity (along $a$- and $c$- axes) from extrapolations of the optical conductivity to zero frequency. The dashed black lines correspond to the theoretical optical conductivities for $a$- and $c$- axis orientations obtained from the first-principles calculations.}
\label{fig1}
\end{figure}

Fig. \ref{fig1}(A) and \ref{fig1}(B) depict the optical conductivity for the $a$- and $c$-axes of the Ta$_{2}$NiSe$_{5}$ sample, respectively. The optical conductivities were obtained from the measured reflectance using the well-known Kramers-Kronig analysis\cite{wooten}. In Fig. \ref{fig1}(A), the optical conductivity of 9 K shows a strong and sharp interband transition (or peak) near 3050 cm$^{-1}$ with an optical gap on the low frequency side of the peak. As the temperature increases, the spectral weight of the peak shifts towards the low frequency region, thereby filling up the optical gap. This temperature-dependent behavior of the peak is similar to a typical signature of an optical gap formation\cite{hwang:2005}. A detailed discussion and analysis on this optical gap and the temperature-dependence of the 3050 cm$^{-1}$ peak will be covered in the following section. This peak centered near 3050 cm$^{-1}$ seems to be closely related to the flat valence band (or the proposed excitonic condensation feature) near the $\Gamma$-point in the Brillouin zone, which was observed via ARPES studies\cite{Wakisaka:2009,Wakisaka:2012,Seki:2014}, since its temperature-dependent behavior and energy scale are similar to those of the flat valence band. The same sharp interband transition has been reported in a recent study of Ta$_{2}$NiSe$_{5}$ probed using spectroscopic ellipsometry\cite{lu:2017,Larkin:2017}. Interestingly, we also observe some more sharp peaks in the optical conductivity (associated with interband transitions) in a higher energy region above 5000 cm$^{-1}$, with these peaks displaying a temperature-dependence behavior similar to that of the 3050 cm$^{-1}$ peak. The similar temperature-dependence behavior will also be discussed later (refer to Fig. \ref{fig2}(B) and adjoining discussion). We have also extracted the dc resistivity from the optical conductivity by extrapolation to $\omega = 0$. The extracted dc resistivity for both $a$- and $c$- axes of Ta$_{2}$NiSe$_{5}$ single crystal is displayed in the inset of Fig. \ref{fig1}(B). The temperature-dependence profile and relative values of the extracted dc resistivity are consistent with those reported ones in recent literature\cite{lu:2017}. The optical conductivity of the Ta$_{2}$NiSe$_{5}$ single crystal along the $c$-axis (depicted in Fig. \ref{fig1}(B)) displays a monotonic temperature-dependence. A strong interband transition peaked near $\sim$15,000 cm$^{-1}$ was observed, with an energy corresponding to the $d$-$d$ transition between the valence Ni-$d$ and conduction Ta-$d$ orbitals\cite{lee:2005}. There was also a significant absorption below this transition, which is not the focus of this article.

First-principles calculations were performed to compute the anisotropic optical properties of Ta$_{2}$NiSe$_{5}$. We adopted the full-potential linearized augmented plane wave (FP-LAPW) implemented in Wien2k\cite{WIEN2K:2001} to calculate the band structure with a number of exchange-correlation functionals, including the generalized gradient approximation (GGA), GGA+U, van der Waals force correction (vdW)\cite{Tkatchenko:2009}, and their hybrid functionals. We obtained insulating ground-states when we used two GGAs: Perdew, Burke, and Ernzerhof (PBE)\cite{Perdew:1996} and modified Becke-Johnson (mBJ)\cite{Tran:2009}. We found that the electronic ground state was semi-metallic when the experimental lattice constants were used\cite{Sunshine:1985}; therefore, we fully relaxed the crystal structure using the PBE and then used the relaxed geometry in further calculations for the band structure and optical conductivity. Note that the volume of our relaxed structure is nearly 14\% larger than the experimental volume. This difference due to the distance between layers ($b$ lattice constant) as shown in the inset of Fig. 1 (A): (i) the relaxed $b$ lattice constant is 14.459 \AA, which is nearly 13\% larger than the experimental one. (ii) the a and c lattice constants are 3.509 \AA (+0.4\%) and 15.732 \AA (+0.6\%), respectively. The relaxed structure was chosen because our calculated electronic properties are similar to previous {\it ab-initio} properties found using the experimentally determined lattice constants\cite{Kaneko:2012}. The reciprocal space integration was approximated by sampling the Brillouin zone with a 28 $\times$6 $\times$ 28 mesh of the Monkhorst-Pack scheme.

The inset in Fig. \ref{fig1}(A) displays the electronic dispersion along the X-$\Gamma$-X direction, which is chosen to compare with the ARPES study results\cite{Wakisaka:2012,Seki:2014}. The Fermi energy ($E_F$) is set to the top of the valence band. The Ni-$d$ and Ta-$d$ orbitals account for the majority of the valence and conduction bands, respectively, as evident from the inset of Fig. \ref{fig1}(A). The results obtained with the mBJ functional were similar to those found with PBE, although the bandgap is twice that obtained with PBE. In general, the results are consistent with those reported by Kaneko {\it et al.}\cite{Kaneko:2012}. We calculated the optical conductivity from the dielectric function obtained using the random phase approximation (RPA)\cite{Draxl:2006} and our first-principles calculation results. The calculated optical conductivity ($\sigma_{1-a}(\omega)$ and $\sigma_{1-c}(\omega)$) for the $a$- and $c$- axes are displayed as dashed black lines in Fig. \ref{fig1}(A) and \ref{fig1}(B), respectively. The difference between the two theoretical conductivity spectra is due to the effect of Ta-NiSe-Ta chains along the $a$-axis\cite{Disalvo:1986,Sunshine:1985}. The calculated $\sigma_{1-c}(\omega)$ agrees reasonably with the measured conductivity of the $c$-axis in its overall shape. However, the calculated $\sigma_{1-a}(\omega)$ shows some discrepancy in the low frequency region below 5000 cm$^{-1}$; the theoretical conductivity shows a higher energy gap and a much smaller spectral weight as compared to the measured one at 9 K. We speculate that this discrepancy occurs due to the non-inclusion of electron-hole interactions in our calculations. Therefore, this result may indicate that the sharp interband transition near 3050 cm$^{-1}$, is closely related to the excitonic excitations. Our observation is in line with a previous report on carbon nanotubes\cite{spataru:2004}, wherein a similar set of calculations (with and without including electron-hole interactions) yielded a similar difference between two results. It is important to note that it is not easy to include electron-hole interactions in the electronic structure calculations for a complex system like Ta$_{2}$NiSe$_{5}$.

\section{Temperature-dependent excitonic insulator gap and interband transitions}

\begin{figure}[!htbp]
  \vspace*{-0.3 cm}%
  \centerline{\includegraphics[width=4.0 in]{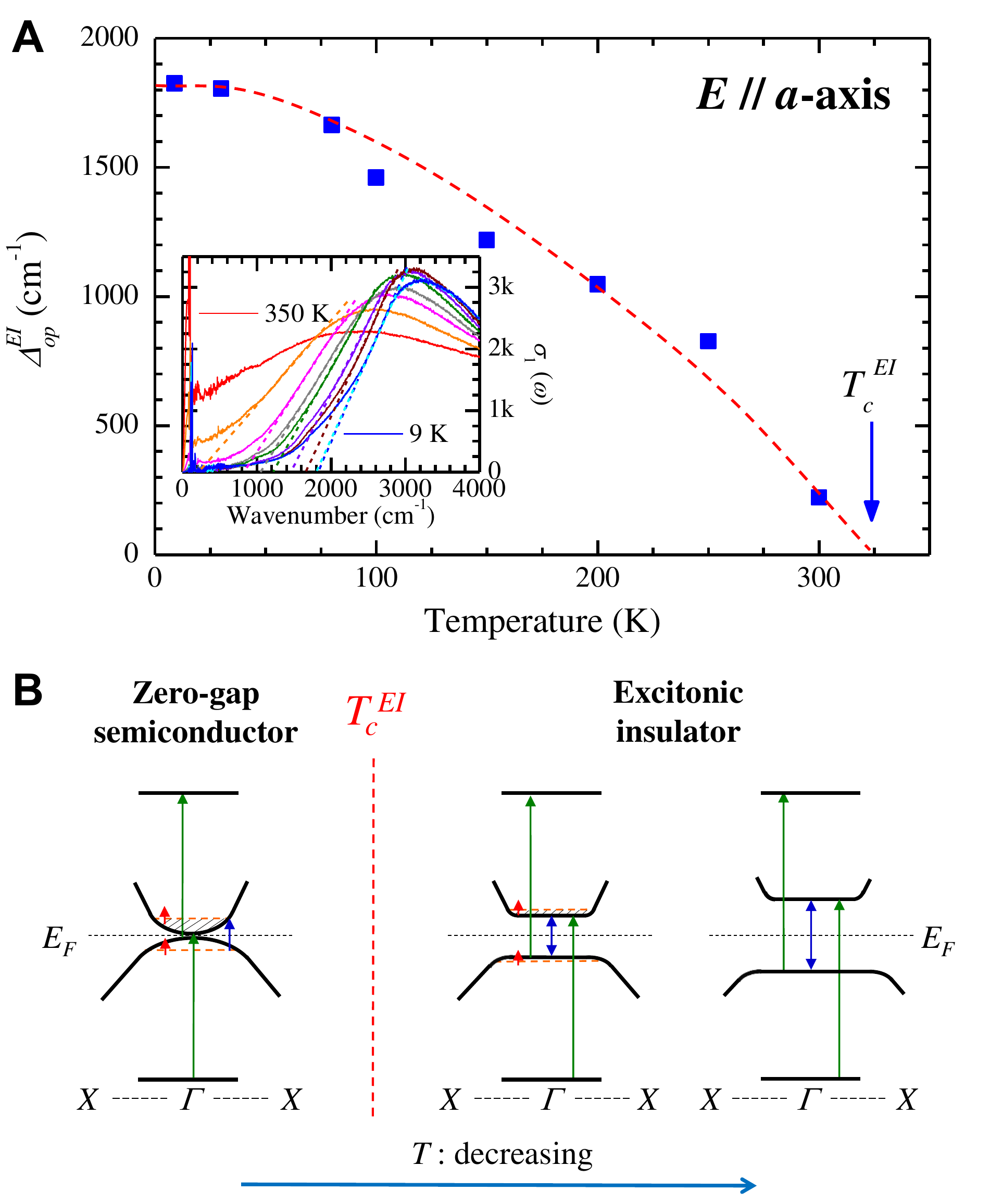}}%
  \vspace*{-0.5 cm}%
\caption{(Color online) Excitonic insulator energy gap and temperature-dependent interband transitions. (A) The temperature-dependent EI gap ($\Delta_{op}^{EI}$) of Ta$_{2}$NiSe$_{5}$ obtained from the optical conductivity. The red dashed line is a guide to the eye. The inset shows the procedure by which the EI gap was obtained. (B) A schematic diagram of the temperature-dependent evolution of the EI gap in Ta$_{2}$NiSe$_{5}$, above and below the transition temperature ($T_c^{EI}$). The intraband transitions are denoted with red arrows. The upward blue arrow denotes the interband transition between the two parabolic bands near the Fermi level. The double direction blue arrows denote the transitions between the two flat bands, which form the EI energy gap. The other interband transitions involving bands near the Fermi level are depicted with green arrows.}
\label{fig2}
\end{figure}

Fig. \ref{fig2}(A) displays the temperature-dependent optical gap in the $a$-axis optical conductivity of single crystal Ta$_{2}$NiSe$_{5}$, which may stem from the formation of excitonic condensation. The inset depicts the method by which the optical gap is extracted from the optical conductivity data; this is an approximate method in order to see a temperature-dependence of the optical gap. We note that there is some amount of spectral weight below the gap, whose origin is not clearly figured out yet. We observe the optical gap starting to open below $\sim$325 K which is the onset temperature, $T_c^{EI}$ marked with an arrow. The extracted gap opening temperature ($T_c^{EI}$) obtained with this approach appears to be almost identical to the structural transition temperature ($T_c^{Str}$). Hence, the two phenomena with the characteristic onset temperatures appear to be closely related to each other. The magnitude of the optical gap increases monotonically with decrease in the temperature and we find that $\Delta_{op}^{EI}(T)/\Delta_0 \simeq 1-[T/T_c^{EI}]^2$ (refer to the following section). The size of the full gap ($\Delta_0$) is $\sim$1800 cm$^{-1}$ (or 0.22 eV) and is consistent with previously reported value ($\sim$0.16 eV at 150 K) in recent literature\cite{lu:2017}, after accounting for the temperature-dependence. For the Ta$_{2}$NiSe$_{5}$ system, this gap can be attributed to the exciton binding energy\cite{lu:2017}, which is much larger than exciton binding energies of bulk (or three-dimensional) semiconductors. The extremely large binding energy can be understood if we consider that the excitons in Ta$_{2}$NiSe$_{5}$ exist along the one-dimensional chains, where long-range Coulomb interaction between an electron and a hole can exist\cite{london:1959,spataru:2004}.

In Fig. \ref{fig2}(B), with the help of schematics, we illustrate the evolution of the electronic structure in the Ta$_{2}$NiSe$_{5}$ sample for temperatures above and below the EI transition temperature ($T_c^{EI}$). Above $T_c^{EI}$, Ta$_{2}$NiSe$_{5}$ is known as a zero-gap semiconductor (ZGSC)\cite{lu:2017}. Therefore, the electronic structure near the Fermi surface can be depicted with two parabolic valence and conduction bands, which are nearly touching each other. Since the temperature is quite large (above $\simeq$ 325 K), we probably have some thermally promoted electrons (or holes) at the bottom of the conduction band (on the top of the valence band), which yields a finite dc conductivity (refer to the red arrows). We expect a broad peak in the optical conductivity at a finite frequency, due to interband transition between the conduction and valence bands near the Fermi level (refer to the blue arrow). We also may have other empty and filled flat bands near the $\Gamma$ point below and above these parabolic bands as shown in the figure and these bands (refer to the black horizontal lines) may not depend on the temperature. These flat bands probably exist along the quasi-one dimensional chains and, in fact, the theoretical calculation shows similar flat bands near the $\Gamma$ point (refer to the inset of Fig. \ref{fig1}(A)). For the ZGSC phase, we expect both intraband and interband transitions from filled states below the Fermi energy to empty states above. These intraband and interband transitions are shown with red arrows, and blue and green arrows, respectively. In fact, we observe these intraband and interband transitions in the measured optical conductivity at 350 K (refer to Fig. \ref{fig1}(A)), which appear as finite dc conductivity and broad peaks, respectively. When the temperature decreases below $T_c^{EI}$, the EI gap ($\Delta^{EI}$) opens and gradually increases, as evident from Fig. \ref{fig2}(A). In this case, the bottom of the valence band and the top of the conduction band are flat and parallel to each other; these flat bands get wider as the temperature is lowered further. Therefore, in the EI phase, a distinct energy gap near the Fermi level, results in a sharp peak just above the gap in the optical conductivity (see in Fig. \ref{fig1}(A)) since the joint density of states may have a singularity. We have other peaks (or interband transitions) in the high frequency region (above 5000 cm$^{-1}$) and observe that these peaks shift to a higher energy and become better defined as the temperature of the sample is lowered. From the schematic, similar temperature-dependence behaviors of all the interband transitions can be rationalized if we consider that the flat bands near the Fermi level induced by the EI phase transition could be involved in the interband transitions in the high frequency region. Our qualitative description of the temperature-dependent behavior of the higher-energy interband transitions is our speculation. It is not completely confirmed by rigorous quantitative analysis yet.

\section{Temperature-dependent accumulated spectral weight and excitonic superfluid plasma frequency}

\begin{figure}[!htbp]
  \vspace*{0.0 cm}%
  \centerline{\includegraphics[width=4.2 in]{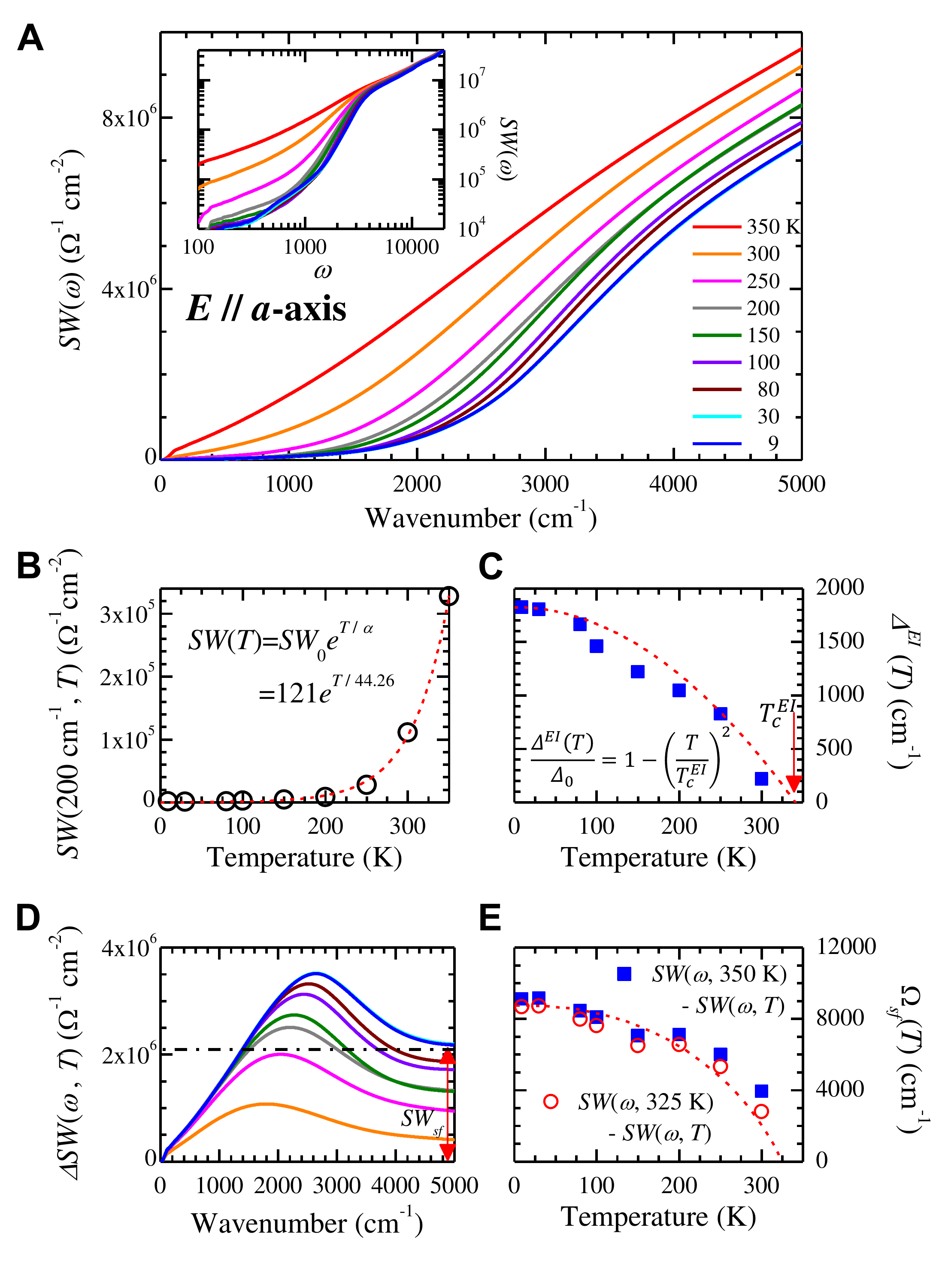}}%
  \vspace*{-0.5 cm}%
\caption{(Color online) Temperature-dependent accumulated spectral weight and excitonic superfluid spectral weight. (A) The accumulated spectral weights of Ta$_{2}$NiSe$_{5}$ along the $a$-axis as a function of frequency at various temperatures. The inset shows the same quantity in log-log scales. (B) The accumulated spectral weight at 200 cm$^{-1}$ as a function of temperature. Red dashed line is an exponential fit to the data. (C) The excitonic energy gap ($\Delta^{EI}(T)$) as a function of temperature obtained considering the thermal effects (see the text for a detailed description) and the excitonic optical energy gap ($\Delta_{op}^{EI}(T)$) extracted from the optical conductivity (refer to Fig. 3(A)). The equation ($\Delta^{EI}(T)/\Delta_0 = 1 - (T/T_c^{EI})^2 $) is for the red dashed line with $T_c^{EI} =$ 338 K. The red arrow indicates $T_c^{EI}$. (D) A quantity, $\Delta SW(\omega, T) \equiv SW(\omega, 350 K) - SW(\omega, T)$. The red arrow shows the missing spectral weight which is closely related to the superfluid spectral weight. (E) Extracted excitonic superfluid plasma frequency ($\Omega_{sf}(T)$) as a function of temperature shown with solid blue squares. The open red circles are adjusted excitonic superfluid plasma frequency. (see in the text) The dashed line is a guide to the eye for the open circles.}
\label{fig3}
\end{figure}

In general, an optical gap formation results in spectral weight redistributions in the optical conductivity. We studied the spectral weight redistribution of the first interband transition peaked near 3050 cm$^{-1}$ with respect to both frequency and temperature. The accumulated spectral weight ($SW$) is a useful physical quantity for studying the spectral weight redistribution and can be defined as $SW(\omega,T) \equiv \int_{0^+}^{\omega}\sigma_1(\omega',T)d\omega'$. In Fig. \ref{fig3}(A), $SW(\omega)$ of the Ta$_{2}$NiSe$_{5}$ sample are displayed at various temperatures in a frequency range up to 5,000 cm$^{-1}$. All the accumulated spectral weights were observed to be more or less parallel to one another above $\sim$4000 cm$^{-1}$ while a small amount of suppression in the accumulated spectral weight occurred below $\sim$4000 cm$^{-1}$, for temperatures below the transition temperature ($T_c^{EI}$). In the inset, $SW(\omega)$ for a wider spectral range up to 20,000 cm$^{-1}$ in log-log scales is displayed.

In Fig. \ref{fig3}(B), $SW(\omega)$ at 200 cm$^{-1}$ as a function of temperature is shown. We chose a low frequency of 200 cm$^{-1}$, well below the full EI gap (1800 cm$^{-1}$) to study only the thermal excitation effect, excluding other absorptions at high frequencies. It is important to note that we have subtracted 10,000 $\Omega^{-1}$cm$^{-2}$ from the obtained $SW(200 \mbox{cm}^{-1}, T)$ to exclude contributions from the low frequency phonons. We show an exponential fit (red dashed line) to the data; the equation employed in the fit is $SW(T) \simeq SW_0 \exp[T/\alpha] = 121 \exp[T/44.26 \:\mbox{K}]$, where $SW_0$ is the spectral weight up to 200 cm$^{-1}$ at $T =$ 0 K. Since this accumulated spectral weight at low frequency is proportional to the charge carrier concentration due to a predominant thermal effect, it can be written as $SW(T) \simeq SW_0 \exp[-\{\Delta^{EI}(T)-\Delta_0\}/T]$, where $\Delta^{EI}(T)$ is the EI gap and $\Delta_0 \equiv \Delta^{EI}(0)$. By combining these two equations, we obtain $\Delta^{EI}(T)/\Delta_0 = 1 - T^2/[\Delta_0 \: \alpha] = 1 - (T/T_c^{EI})^2$ ($\geq 0$), where $T_c^{EI}$ ($= \sqrt{\Delta_0\alpha}$) is the onset temperature of the gap. If we consider $\Delta_0$ as the full optical EI gap ($\simeq$ 1800 cm$^{-1}$) and $\alpha =$ 44.26 (from the exponential fit), then the onset temperature ($T_c^{EI}$) is 338 K. In Fig. \ref{fig3}(C), we plot the EI gap, $\Delta^{EI}(T)$, with the red dashed line along with the optical excitonic gap, $\Delta_{op}^{EI}(T)$, which was obtained directly from the optical conductivity (refer to Fig. \ref{fig2}(A)). From the figure, it is evident that these two results are in good agreement with each other.

Furthermore, from the accumulated spectral weight, we calculate an interesting physical quantity, the superfluid plasma frequency of the excitonic condensation. In Fig. \ref{fig3}(D), we present a differential quantity, $\Delta SW(\omega, T) \equiv SW(\omega, 350 K) - SW(\omega, T)$ at various temperatures. This quantity seems to consist of two components: one is unrecovered spectral weight near 5000 cm$^{-1}$ (or missing spectral weight) due to the excitonic condensation marked with the red arrow and the other is the peak near 2600 cm$^{-1}$ due to the thermal broadening effects, which may come from two different temperatures of the two different phases (here, the most pronounced peak between 9 K and 350 K). We note that the missing spectral weight is not recovered up to 20,000 cm$^{-1}$ and above. We present a more detailed discussion on this quantity ($\Delta SW(\omega, T)$), comparing it with that of the superconductors, in the following section. The superfluid spectral weight ($SW_{sf}$) can be related to the superfluid plasma frequency ($\Omega_{sf}$) as $\Omega_{sf}(T) \equiv \sqrt{(120/ \pi) \: SW_{sf}(T)}$. We note that the numerical factor $\pi/120$ is the unit conversion factor; here $\Omega_{sf}$ and $SW_{sf}$ are in cm$^{-1}$ and $\Omega^{-1}$cm$^{-2}$ units, respectively. We display the superfluid plasma frequency (blue solid squares) as a function of temperature in Fig. \ref{fig3}(E). $\Omega_{sf}$ is gradually decreasing with increasing the temperature and then eventually going to zero near the EI onset temperature, $T_c^{EI}$. Therefore, the onset temperature of the superfluid condensation seems to be the same as that of the EI gap. It is worthwhile to note that up to now we used the spectral weight at 350 K as the reference spectral weight for getting the differential spectral weight at various temperatures since we do not have data closer to the transition temperature. If we use a linear interpolated spectral weight (at 325 K) between 300 K and 350 K as the reference spectral weight we will have slightly lower excitonic condensation plasma frequencies than the values obtained using the spectral weight at 350 K as the reference spectral weight, as displayed in Fig. \ref{fig3}(E) with red open circles.

\section{Discussion: Excitonic insulators and superconductors}

\begin{figure}[!htbp]
  \vspace*{-0.5 cm}%
  \centerline{\includegraphics[width=6.0 in]{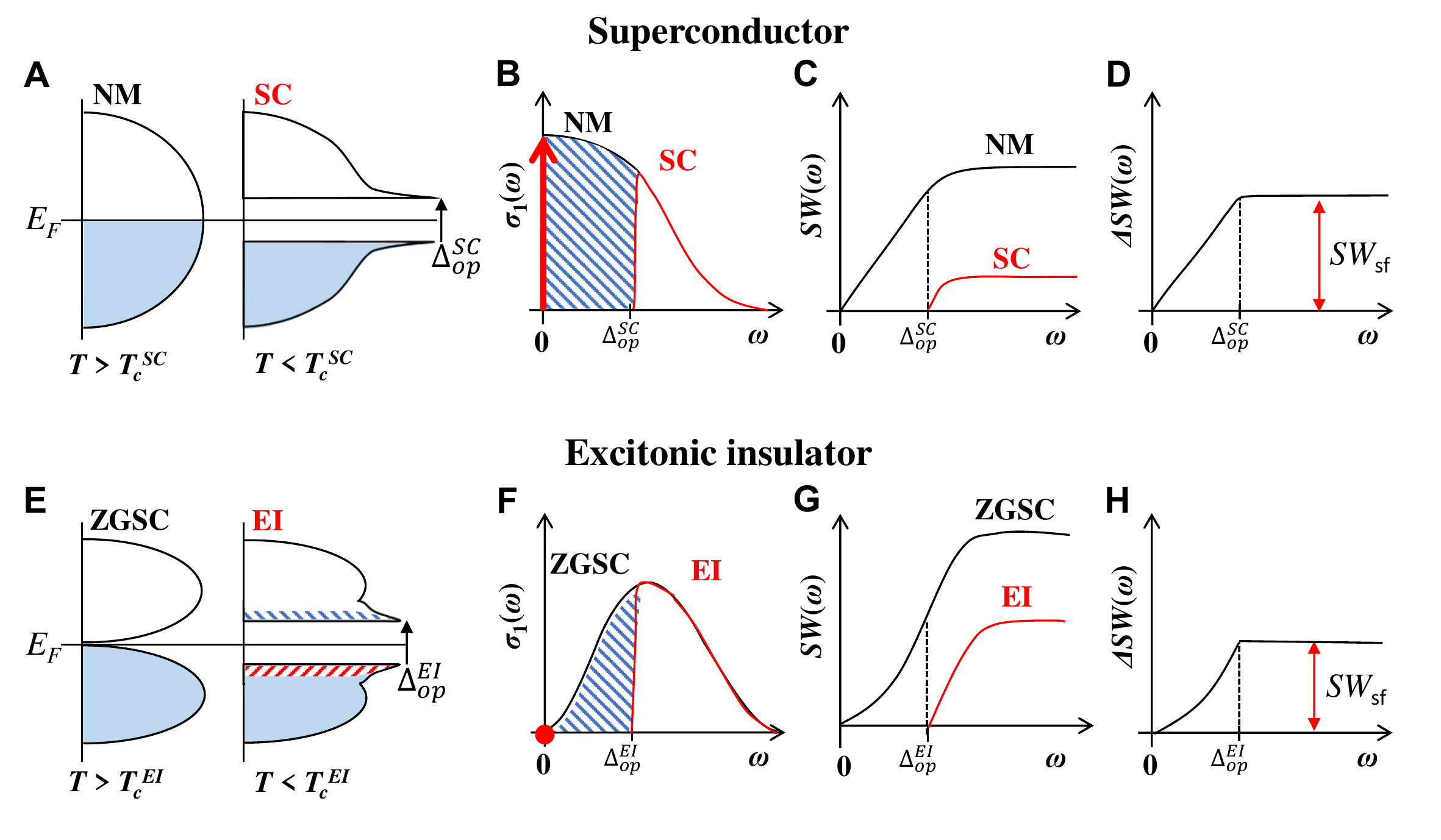}}%
  \vspace*{-0.5 cm}%
\caption{(Color online) Comparison of excitonic insulators with superconductors. (A) The densities of states (DOS's) of normal metal (NM) and superconductor (SC). (B) The corresponding optical conductivities ($\sigma_1(\omega)$). The thick red vertical arrow indicates the superfluid spectral weight condensed at zero frequency. (C) The accumulated spectral weights ($SW(\omega)$) of NM and SC. (D) The differential spectral weight ($\Delta SW(\omega)$) between NM and SC. (E) DOS's of zero-gap semiconductor (ZGSC) and excitonic insulator (EI); the red (electrons) and blue (holes) dashed horizontal lines show the thermal excitations. (F) $\sigma_1(\omega)$ of ZGSC and EI; the black (for ZGSC) and red (for EI). The red dot at zero frequency is the excitonic superfluid with zero spectral weight. (G) $SW(\omega)$ of ZGSC and EI. (H) $\Delta SW(\omega)$ of ZGSC and EI. The red arrow shows the missing spectral weight, which is closely related to the excitonic superfluid plasma frequency.}
\label{fig4}
\end{figure}

It is worthwhile to compare the condensation in the EI with that in a superconductor. In Fig. \ref{fig4}(A) - (H), we compare the two material systems: $s$-wave superconductors (SC) and EI schematically. We depict the transition from a normal metal (NM) to a SC and a zero-gap semiconductor (ZGSC) to an EI, as the temperature is lowered from above to below the transition temperatures ($T_c^{SC}$ and $T_c^{EI}$), respectively, through spectral weight redistributions. Fig. \ref{fig4}(A) shows the density of states (DOS's) of the normal metal and the superconductor, \ref{fig4}(B) shows the corresponding optical conductivities ($\sigma_1(\omega)$), \ref{fig4}(C) shows the accumulated spectral weights ($SW(\omega)$), and \ref{fig4}(D) shows the differential spectral weights ($\Delta SW(\omega) \equiv SW(\omega, T>T_c^{SC})-SW(\omega, T<T_c^{SC})$). In the superconductor, the superfluid spectral weight of condensed Cooper (or electron-electron) pairs appears as a delta function at zero frequency, which is marked with a thick red vertical arrow in Fig \ref{fig4}(B); the hatched area below the superconducting gap ($\Delta_{op}^{SC}$) seems to have disappeared\cite{glover:1956}. Therefore, this area is termed as the missing spectral weight. At high frequencies well above the SC gap, the accumulated spectral weights of NM and SC will differ by the missing spectral weight, as shown in Fig. \ref{fig4}(C). When the condensed Cooper pairs are broken by thermal or other processes, the missing spectral weight reappears back in a finite frequency region. The differential spectral weight ($\Delta SW(\omega)$) in the high frequency region will clearly show the missing spectral weight, as depicted in Fig. \ref{fig4}(D).

We also sketch the corresponding four physical quantities (DOS's, $\sigma_1(\omega)$, $SW(\omega)$, and $\Delta SW(\omega)$) for both the ZGSC and the EI in Fig. \ref{fig4}(E) - (H). In the EI, there is a superfluid spectral weight associated with the condensed electron-hole pairs; however, the neutral excitonic superfluid will be located at zero frequency with zero spectral weight since the neural excitons cannot contribute to the electrical conductivity\cite{kozlov:1965}. The electrons involved in the condensation will be disappeared in the EI states as in the SC state; this also causes a missing spectral weight. However, the electrons in the EI state appear nowhere in the whole frequency range while the electrons in the SC state will appear at zero frequency as a delta function. Therefore, in the EI state the optical sum rule may be violated; the missing spectral weight (or missing electron density) still remains in the sample but is just optically invisible. The EI has singularities at the bottom of the conduction band and on the top of the valence band, similar to the superconductor, as evident in Fig. \ref{fig4}(E). These singularities are a signature of the condensation, which leads to the flat valence band in ARPES dispersion and a sharp peak in the optical conductivity as shown in Fig. \ref{fig1}(A) and sketched in Fig. \ref{fig4}(F). The singularity at the bottom of the conduction band consists of the electron states while the singularity on the top of the valence band consists of the hole states (illustrated with two different colors in Fig. \ref{fig4}(E)). Here we assume that both ZGSC and EI phases are at the absolute zero temperature. The resulting accumulated spectral weight in Fig. \ref{fig4}(G) looks analogous to what was observed in our measured accumulated spectral weights in Fig. \ref{fig3}(A), for frequencies below 3500 cm$^{-1}$. However, they appear to be different in the high frequency region well above the EI gap. The accumulated spectral weights of ZGSC and EI flatten and are parallel to each other at the high frequency region while the measured ones are still parallel to each other but keep increasing. This difference at the high frequency region can be explained by considering other interband absorptions existing at higher frequencies. If we include the interband absorption bands at higher frequencies, the results will show the continuous parallel increase at high frequencies, as in the measured data (Fig. \ref{fig3}(A)). In Fig. \ref{fig4}(H) we display the differential spectral weight ($\Delta SW(\omega)$). We do not see the peak, which was observed in the experimental differential spectral weight (refer to Fig. \ref{fig3}(D)). If we take the thermal broadening effects, which may come from the temperature difference between ZGSC and EI phases, into account we will have the peak in the $\Delta SW(\omega)$.

\section{Conclusion}

In conclusion, we observed a strong and sharp peak around 3050 cm$^{-1}$ in the optical conductivity (along the $a$-axis) of Ta$_{2}$NiSe$_{5}$ at low temperatures along with a well-defined optical gap on the low frequency side of the peak. This peak corresponds to an interband transition, which shows a characteristic strong temperature-dependence, a behavior previously attributed to that of the flat valence band in observations made of Ta$_{2}$NiSe$_{5}$ with ARPES\cite{Wakisaka:2009,Wakisaka:2012,Seki:2014}. The results of our first-principles calculations were in good agreement with the overall experimental optical data for both $a$- and $c$-axes except for the strong and sharp peak in the $a$-axis conductivity. We speculate that this discrepancy between experiment and theory arises from the fact that the electron-hole interactions in Ta$_{2}$NiSe$_{5}$ were not included in the theoretical calculations. This result probably indicates that the strong and sharp peak results from the electron-hole interactions for this system. Furthermore, the spectral weight redistribution analysis demonstrates that the excitonic condensation of electron-hole pairs can occur below the onset temperature of the optical gap and the temperature-dependent excitonic superfluid plasma frequency can be obtained from the measured optical data. These interesting findings illustrate the new opportunities for further investigations on Ta$_{2}$NiSe$_{5}$, and other excitonic insulators.

%
%
\acknowledgments JH acknowledges financial support from the National Research Foundation of Korea (NRFK Grant No. 2017R1A2B4007387). Y.S.S acknowledges financial support from the National Research Foundation of Korea (NRFK Grant No. 2016R1A6A3A11933016). M.J.E and J.S.K acknowledge support from the NRF through SRC (Grant No. 2011-0030785) and the Max Planck POSTECH/KOREA Research Initiative (Grant No. 2011-0031558) Programs, and also from IBS (No. IBSR014-D1-2014-a02). J.H. gratefully acknowledges Yunkyu Bang and Sung-Sik Lee for helpful discussions.

\bibliographystyle{apsrev4-1}
\bibliography{TNS215_v6}

\end{document}